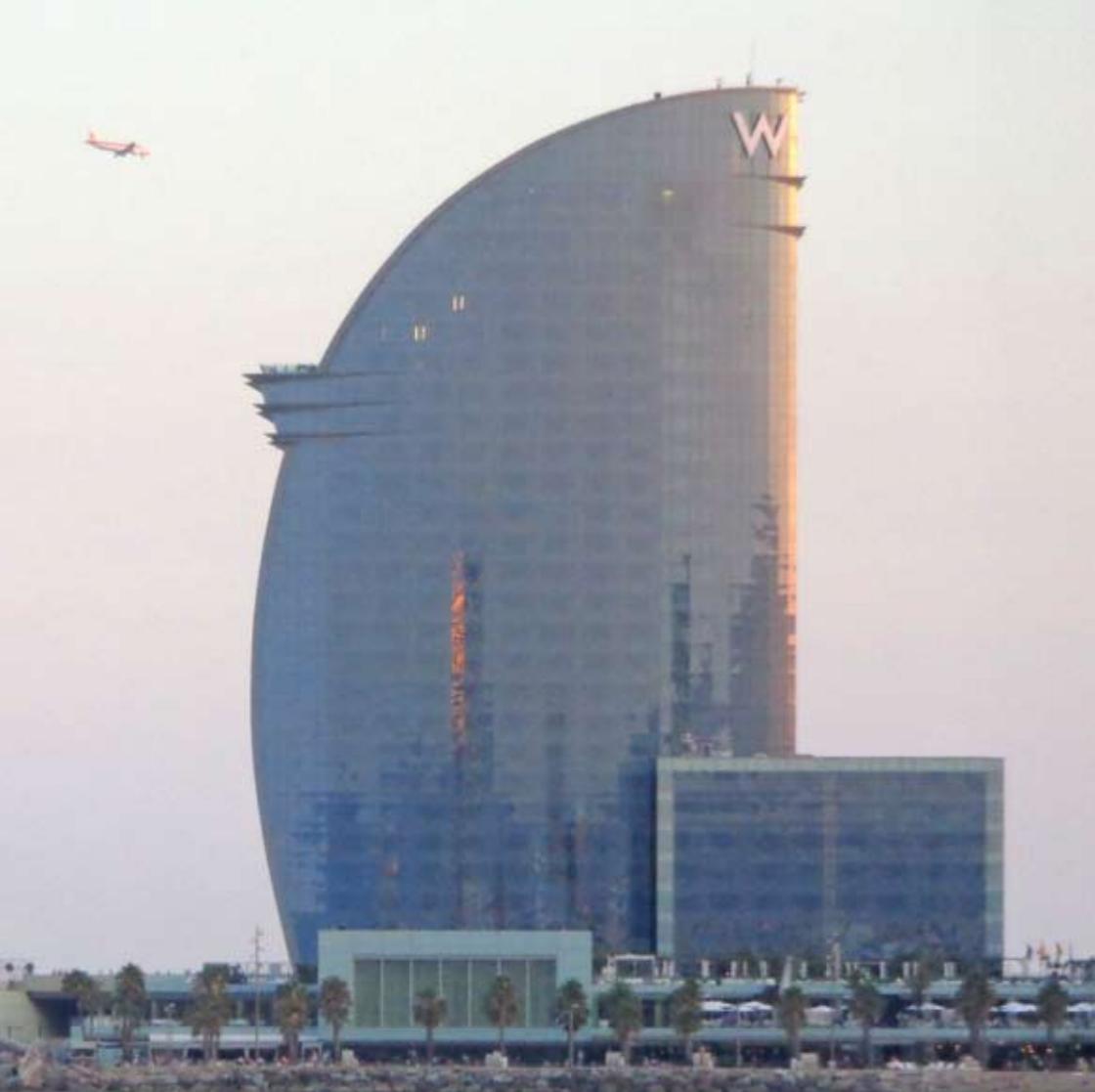

# Proceedings of
# VI International Workshop on Locational Analysis and Related Problems

Barcelona, Spain - November 25-27, 2015

# Program Overview

| | Wednesday Nov. 25th | Thursday Nov. 26th | Friday Nov. 27th |
|---|---|---|---|
| 9:00-10:40 | | SESSION 2: CONTINUOUS LOCATION | SESSION 6: DISCRETE LOCATION |
| 10:40-11:10 | | Coffee break | Coffee break |
| 11:10-12:30 | | Invited Speaker: Matteo Fischetti | Invited Speaker: M. Grazia Speranza |
| 12:30-13:45 | | SESSION 3: LOCATION ON NETWORKS | SESSION 7: ROUTING PROBLEMS |
| 13:45-14:30 | | LUNCH | Locat. Network Meeting |
| 14:30-15:30 | | | LUNCH |
| 15:30-17:10 | | SESSION 4: DISCRETE LOCATION | |
| 17:10-17:40 | REGISTRATION | Coffee break | |
| 17:40-18:10 | OPENING SESSION | SESSION 5: LOCATION ON NETWORKS | |
| 18:10-18:55 | SESSION 1: DISCRETE LOCATION | | |
| 18:55-19:50 | | | |
| 20:00 | Welcome Reception | DINNER | |

# PROCEEDINGS OF THE VI INTERNATIONAL WORKSHOP ON LOCATINAL ANALYSIS AND RELATED PROBLEMS (2015)

# Preface

The International Workshop on Locational Analysis and Related Problems will take place during November 25-27, 2015 in Barcelona (Spain). It is organized by the Spanish Location Network and Location Group GELOCA (SEIO). GELOCA is a working group on location belonging to the Statistics and Operations Research Spanish Society. The Spanish Location Network is a group of more than 140 researchers distributed into 16 nodes corresponding to several Spanish universities. The Network has been funded by the Spanish Government.

Every year, the Network organizes a meeting to promote the communication among its members and between them and other researchers, and to contribute to the development of the location field and related problems. Previous meetings took place in Sevilla (October 1-3, 2014), Torremolinos (Málaga, June 19-21, 2013), Granada (May 10-12, 2012), Las Palmas de Gran Canaria (February 2-5, 2011) and Sevilla (February 1-3, 2010).

The topics of interest are location analysis and related problems. This includes location, routing, networks, transportation and logistics models; exact and heuristic solution methods, and computational geometry, among others.

The organizing committee.

**Scientific committee:**

- Emilio Carrizosa (U. de Sevilla)
- Ángel Corberán (U. de Valencia)
- Elena Fernández Aréizaga (U. Politécnica de Cataluña)
- Alfredo Marín Pérez (U. de Murcia)
- Juan A. Mesa (U. de Sevilla)
- Blas Pelegrín (U. de Murcia)
- Justo Puerto Albandoz (U. de Sevilla)
- Antonio M. Rodríguez-Chía (U. de Cádiz)

**Organizing committee:**

- Maria Albareda-Sambola (U. Politécnica de Cataluña)
- Luisa Martínez-Merino (U. de Cádiz)
- Francisco A. Ortega (U. de Sevilla)
- Dolores R. Santos Peñate (U. de Las Palmas de G.C)

# Contents

PROGRAM

# Wednesday November 25th

## 17:10-17:40 Registration

## 17:40-18:10 Opening Session

## 18:10-19:50 Session 1: Discrete Location

> Locating capacitated unreliable facilities
> M. Albareda-Sambola, M. Landete, J.F. Monge, and J.L. Sainz-Pardo

> Conditions to LP relax the allocation variables of the reliability fixed-charge location problem
> J. Alcaraz, M. Landete, J.F. Monge and J.L. Sainz-Pardo

> The maximal covering location bi-level problem
> J.F. Camacho-Vallejo, M.S. Casas-Ramírez, J.A. Díaz and D.E. Luna

> Using an interior-point method for huge capacitated multiperiod facility location
> J. Castro, S. Nasini and F. Saldanha-da-Gama

## 20:00 Welcome Reception

# Thursday November 26th

## 9:00-10:40 Session 2: Continuous Location

Multisource linear regression
V. Blanco and J. Puerto

A dispersion model in ODEs
R. Blanquero, E. Carrizosa, M.A. Jiménez-Cordero and B. G. Tóth

Ordered Weighted Average Optimization in multiobjective spanning tree problems
E. Fernández, M. A. Pozo, J. Puerto and A. Scozzari

Optimal routes to a destination for airline expansion
B. Pelegrín Pelegrín, P. Fernández Hernández, and J.D. Pelegrín García

## 10:40-11:10 Coffee break

## 11:10-12:30 Invited Speaker: Matteo Fischetti

Simplified Benders cuts for facility location
M. Fischetti, I. Ljubić and M. Sinnl

## 12:30-13:45 Session 3: Location on Networks

Rapid transit network design: competition and transfers
L. Cadarso and Á. Marín

A decomposition scheme for the railway rapid transit depot location and rolling stock circulation problem
D. Canca, A. de los Santos, E. Barrena and G. Laporte

Location of emergency units in collective transportation line Networks
T. Tan and J.A. Mesa

## 13:45-15:30 Lunch

## 15:30-17:10 Session 4: Discrete Location

The p-center problem with uncertainty in the demands
M. Albareda-Sambola, L.I. Martínez-Merino and A.M. Rodríguez-Chía

Set-packing problems in discrete location
A. Marín and M. Pelegrín

Convex ordered median problem
D. Ponce and J. Puerto

A variable neighborhood search approach for the 3-maneuver aircraft conflict resolution problem
A. Alonso-Ayuso, L.F. Escudero, F.J. Martín-Campo, N. Mladenović

## 17:10-17:40 Coffee break

## 17:40-18:55 Session 5: Location on Networks

Robust rapid transit railway rescheduling
Á. Marín and L. Cadarso

Dial-a-ride problems in presence of transshipments
J.A. Mesa, F. A. Ortega and M.A. Pozo

Locating new stations and road links on a road-rail network
F. Perea

## 20:00 Dinner

# Friday November 28th

## 9:00-10:40 Session 6: Discrete Location

On $k$-centrum optimization with applications to the location of extensive facilities on graphs and the like
J. Puerto and A.M. Rodríguez-Chía

Robust $p$-median problem with vector autoregressive demands
E. Carrizosa, A.V. Olivares-Nadal and P. Ramírez-Cobo

New products supply chains: the effect of short lifecycles on the supply chain network design
M.B.C. Menezes. K. Luo and O. Allal-Cherif

Analyzing the impact of capacity volatility on the design of a supply chain network
D. Ruiz-Hernández, M.B.C. Menezes and S. Gueye

## 10:40-11:10 Coffee Break

## 11:10-12:30 Invited Speaker: M. Grazia Speranza

Kernel search for location problems
M. G. Speranza

## 12:30-13:45 Session 7: Routing Problems

The probabilistic pickup and delivery problem
E. Benavent, M. Landete, J.J. Salazar and G. Tirado

On the hierarchical rural postman problem on a mixed graph
M. Colombi, Á. Corberán, R. Mansini, I. Plana and J.M. Sanchis

Strategic oscillation for a hub location problem with modular link capacities
Á. Corberán, J. Peiró, F. Glover and R. Martín

## 13:45-14:30 Location Network Meeting

## 14:30-15:30 Lunch

# INVITED SPEAKERS

# Simplified Benders cuts for facility location

Matteo Fischetti,[1] Ivana Ljubić,[2] and Markus Sinnl[3]

[1]*Dept. of Information Engineering, Univ. of Padua, Italy,*   matteo.fischetti@unipd.it

[2]*ESSEC Business School of Paris, France,*   ljubic@essec.edu

[3]*Dept. of Statistics and Op. Res., Univ. of Vienna, Austria,*   markus.sinnl@univie.ac.at

A simple reformulation of generalized Benders cuts is presented, that greatly simplifies their practical implementation. Successful applications to Facility Location Problems (FLPs) with convex (linear and quadratic) costs will be discussed.

Consider the convex Mixed-Integer (possibly Nonlinear) Problem

$$\min\{\, f(x,y) :\ g(x,y) \leq 0, Ay \leq b,\ y \text{ integer}\,\} \tag{1}$$

where $x \in \Re^m$, $y \in \Re^n$, and functions $f : \Re^{m+n} \mapsto \Re$ and $g : \Re^{m+n} \mapsto \Re^p$ are assumed to be differentiable and convex. In FLPs, $y$ variables are typically associated to facilities, and $x$ variables to allocation decisions.

To simplify our treatment, we assume that $S := \{y : Ay \leq b\}$ is a nonempty polytope, while the convex sets $X(y) := \{x : g(x,y) \leq 0\}$ are nonempty, closed and bounded for all $y \in S$, as it happens for FLPs. Problem (1) can trivially be restated as the *master problem* in the $y$ space

$$\min\{\, w :\ w \geq \Phi(y),\ Ay \leq b,\ y \text{ integer}\,\} \tag{2}$$

where

$$\Phi(y) := \min_{x \in X(y)} f(x,y)$$

is the convex (nonlinear) function expressing the optimal solution value of the problem (1) as a function of $y \in S$, and $w$ is a continuous variable that captures its value in the objective function.

Master problem (2) can be solved by an LP-based branch-and-cut approach where $\Phi(y)$ is under-approximated by linear cuts to be generated on the fly and added to the current LP relaxation. A crucial point is the efficient generation of the approximation cuts. To this end, consider a (possibly noninteger) solution $y^*$ of the LP relaxation of the current master. Because of convexity, $\Phi(y)$ can be underestimated by a supporting hyperplane at $y^*$, so we can write the following generalized Benders (linear) cut

$$w \geq \Phi(y) \geq \Phi(y^*) + \xi(y^*)^T(y - y^*) \tag{3}$$

Here $\xi(y^*)$ denotes a subgradient of $\Phi$ in $y^*$ that can be computed as

$$\xi(y^*) = \nabla_y f(x^*, y^*) + u^* \nabla_y g(x^*, y^*) \tag{4}$$

where $x^*$ and $u^*$ are optimal primal and (Lagrangian) dual solutions of the convex problem obtained from (1) by replacing $y$ with the given $y^*$ [1, 2].

The above formula involves the computation of partial derivatives of $f$ and $g$ with respect to the $y_j$'s, so it is problem specific and sometimes cumbersome to apply. We next introduce a very simple reformulation that makes its implementation straightforward. For a given $y \in S$, $\Phi(y)$ can be computed by solving the convex *slave problem*

$$\Phi(y) = \min\{ f(x, q) : g(x, q) \leq 0, \ y - q = 0 \} \tag{5}$$

The variable-fixing equation in (5) is meant to be imposed as $y \leq q \leq y$ by just modifying the lower and upper bounds on the $q$ variables, so it can be handled very efficiently by the solver in a preprocessing phase when $y$ is given.

By construction, $y$ only appears in the variable-fixing equation in (5), hence the subgradient (4) is just $\xi(y^*) = r^*$, where $r^*$ is the vector of optimal reduced costs returned by a convex solver applied to (5) for $y = y^*$. This leads to completely general and easily computable generalized Benders cut

$$w \geq \Phi(y^*) + \sum_{j=1}^{n} r_j^*(y_j - y_j^*) \tag{6}$$

Extensive computational results for various FLPs confirm the practical effectiveness of the above family of cuts.

# Kernel search for location problems

M. Grazia Speranza [1]

[1]*Department of Economics and Business, University of Brescia, Italy,*
grazia.speranza@unibs.it

In this talk a general heuristic approach, known as Kernel Search, will be presented that has been successfully applied to several MILP problems, and in particular to location problems. The Kernel Search is a general and simple heuristic framework that has been introduced in [1] and [2] for the solution of MILP problems with binary variables, in particular of a portfolio optimization problem and the Multidimensional Knapsack Problem. The original idea was to consider all the problem variables through the solution of a sequence of MILP problems, each restricted to a subset of variables. This restriction is equivalent to setting to 0 a subset of variables. In the sequence of restricted MILP problems the size of the solved problems was increasing because at each iteration new variables may be added to the subset and none was removed. Some enhancements of the Kernel Search were proposed in Guastaroba and Speranza [3], where the removal of variables was allowed. A new variant was also proposed that refines the solution found through the first sequence of restricted problems. The idea of the variant is to perform variable fixing (binary variables set to 1) in the original MILP problem on the basis of the solutions found in the first sequence, and to solve to optimality the MILP problem restricted to the remaining variables.

The Kernel Search has been later extended and applied to two well studied location problems, namely the Capacitated Facility Location Problem (CFLP) and the Single Source Capacitated Facility Location Problem (SS-CFLP). In [4] the Kernel Search has been extended to solve problems where a large number of continuous variables are associated with each binary variable, in particular to the CFLP, while in [5] it has been extended to solve any problem with binary variables, in particular the SSCFLP. The computational results prove that the Kernel Search can solve instances of large size of the CFLP and of the SSCFLP with negligible errors with respect to the optimum, and outperforms previous heuristics.

# ABSTRACTS

# Locating capacitated unreliable facilities [*]

M. Albareda-Sambola,[1] M. Landete,[2] J.F. Monge,[2] and J.L. Sainz-Pardo[2]

[1]*Univ. Politècnica de Catalunya, Barcelona Tech., Terrassa, Spain*  maria.albareda@upc.edu

[2] *Universidad Miguel Hernández, Elche, Spain*  landete@umh.es, monge@goumh.umh.es, jose.sainz-pardo@umh.es

This work is focused on a fixed charge facility location problem where facilities are considered to be capacitated and unreliable. We explore different strategies for considering facility capacities in a reasonably flexible way: facility capacities refer to the maximum acceptable workload in regular conditions, but extra load can be occasionally dealt with in emergency situations where some failures have affected the system. However, these eventual overloads must be kept reasonably small and unlikely to occur. We propose several models, and analyze their capability to generate solutions with these characteristics in a series of computational experiments.

## 1.     Problem definition

The problems we consider are defined on a discrete setting. A set of potential facility locations is given, each with an associated fixed opening cost and a capacity. This set of facilities is partitioned into two subsets. The failing and the non-failing facilities. A failure probability, common to all the failing facilities, is also known. Moreover, it is assumed that facility failures take place independently of each other. On the other hand, a set of customers with different associated demands is considered together with their service costs (proportional to distances) from each of the potential facility locations. Additionally, we consider a cost associated with loosing or outsourcing a customer. In order to introduce this possibility in our models, we just add to the set of potential facilities a dummy non-failing facility

[*]Research funded by the Spanish Ministry of Economy and Competitiveness and ERDF funds through Project MTM2012-36163-C06

with zero fixed opening cost and infinite capacity whose distance to all the customers equals this outsourcing cost.

The goal is to decide on the set of facilities to open and design a distribution pattern such that capacity constraints are *nearly always satisfied*, at a minimum total cost. In this work we consider this total cost to be a convex combination of the sum of fixed and service costs in the scenario where no failures occur, and the expected service cost over all scenarios.

## 2. Modeling assumptions

When considering location problems with unreliable facilities, the system behavior upon failure has to be clearly defined. Do customers have complete information on facility failures? Can customers be always served by their closest available facility? To what extent can the *a priori* assignments be modified to adapt to scenarios with failures?

Since facilities are capacitated and have fixed opening costs, it would be quite expensive to design systems capable to satisfy the capacity constraints even if customers are always served by their closest available facility, as it is assumed in [3]. Although this may be a reasonable assumption to make in the location of essential services, it can be rather unrealistic in more general ones. So, we assume customers can be served by any available facility. However, to keep the effect of failures as local as possible, we consider that customers cannot be all freely relocated at each scenario, as it is assumed in [2], but assignments at different levels are defined *a priori*, like in uncapacitated models such as [1], and each customer is served from the available facility it has been assigned to at the lowest level.

Three model types are proposed: limiting the expected loads, limiting estimates of the expected overloads, or using auxiliary staggered capacities, being the two last strategies the ones yielding the best solutions.

# The $p$-center problem with uncertainty in the demands [*]

Maria Albareda-Sambola,[1] Luisa I. Martínez-Merino,[2] and Antonio M. Rodríguez-Chía,[3]

[1]*Departamento de Estadística e Investigación Operativa, Univ. Politècnica de Catalunya. BarcelonaTech, Barcelona, Spain*,  maria.albareda@upc.edu

[2]*Departamento de Estadística e Investigación Operativa, Universidad de Cádiz, Spain*, luisa.martinezmerino@alum.uca.es

[3]*Departamento de Estadística e Investigación Operativa, Universidad de Cádiz, Spain*,  antonio.rodriguezchia@uca.es

**Keywords:** *p-center problem, probabilistic*

This work deals with the p-center problem, where the aim is to minimize the maximum distance between any user and his center taking into account that the demand occurs in any site with a specific probability. The problem is of interest when locating emergency centers . We consider different formulations for the problem and extensive computational tests are reported, showing the potentials and limits of each formulation on several types of instances. Finally, different techniques to obtain accurate bounds on the optimal solution of the problem are explained.

## 1. Introduction to the problem

The $p$-center problem (pCP) is a well-known discrete optimization location problem which consists of locating $p$ centers out of $n$ sites and assigning (allocating) the remaining $n-p$ sites to the centers so as to minimize the maximum distance (cost) between a site and the corresponding center, see [1, 2]. It was shown in [2] that pCP is NP-hard.

[*]Thanks to Spanish Research projects: MTM2012-36163-C06-05 and MTM2013-46962-C2-2-P

A straight application of the pCP is the location of emergency services like ambulances, hospitals or fire stations, since the whole population should be inside a small radius around some emergency center. pCP has been extensively studied, and both exact and heuristic algorithms have been proposed.

The uncertainties can be generally classified into three different categories: provider-side uncertainty, receiver-side uncertainty, and in-between uncertainty. The provider-side uncertainty may capture the randomness in facility capacity and the reliability of facilities, etc.; the receiver-side uncertainty can be the randomness in demands; and the in-between uncertainty may be represented by the random travel time, transportation cost, etc.

In this paper we focus on the receiver-side uncertainty and in the stochastic programming approach. The SP approach has been widely applied to emergency logistics for short-notice disasters (e.g., hurricanes, flooding, and wild fires) by assuming that possible impacts of these disasters can be estimated based on historical and meteorological data. The common goal of these stochastic location models is to optimize the expected value of a given objective function. A classical example of applying SP to disaster relief is the scenario-based, two-stage stochastic model proposed by [3], for medical supply pre-positioning and distribution in emergency management.

Different formulations of this problem are proposed and we show their strengths and limits. Besides we adapt Variable Neighborhood Search (VNS) which is a heuristic to solve combinatorial problems originally proposed for the $p$-median problem. Computational tests show that VNS provides high quality solutions.

# Conditions to LP relax the allocation variables of the reliability fixed-charge location problem

Javier Alcaraz, Mercedes Landete, Juan F. Monge
and José L. Sainz-Pardo [1]

[1]*Departamento de Estadística, Matemáticas e Informática,*
*Centro de Investigación Operativa,*
*Universidad Miguel Hernández de Elche, Spain,*
*jalcaraz@umh.es, landete@umh.es, monge@umh.es, jose.sainz-pardo@umh.es*

The Reliability Fixed-Charge Location Problem is an extension of the Simple Plant Location Problem which considers that some facilities have a probability of failure. Snyder and Daskin (2005) proposed a binary formulation for the Reliability FixedCharge Location Problem. They defined the two classical families of binary variables on location problems, the location variables and the allocation variables and assumed that the LP relaxation of the allocation variables was straightforward. In this work we show that although for the instances tested in Snyder and Daskin (2005) the assumption was true, the allocation variables can be LP relaxed only if certain conditions hold.

## 1. On LP relaxing the allocation variables

We prove that we can LP relax all the allocation variables associated to non-failable facilities or all the allocation variables associated to failable facilities but not both simultaneously. We also prove that all the allocation variables could be LP relaxed by adding a family of valid equalities to the set of constraints. Finally, we show and demonstrate in which cases we can to LP relax all the allocation variables simultaneously.

## 2. Computational experience

The figure 1 shows the magnitude of the objective value errors in order to illustrate how wrong it is to LP relax all the allocation variables.

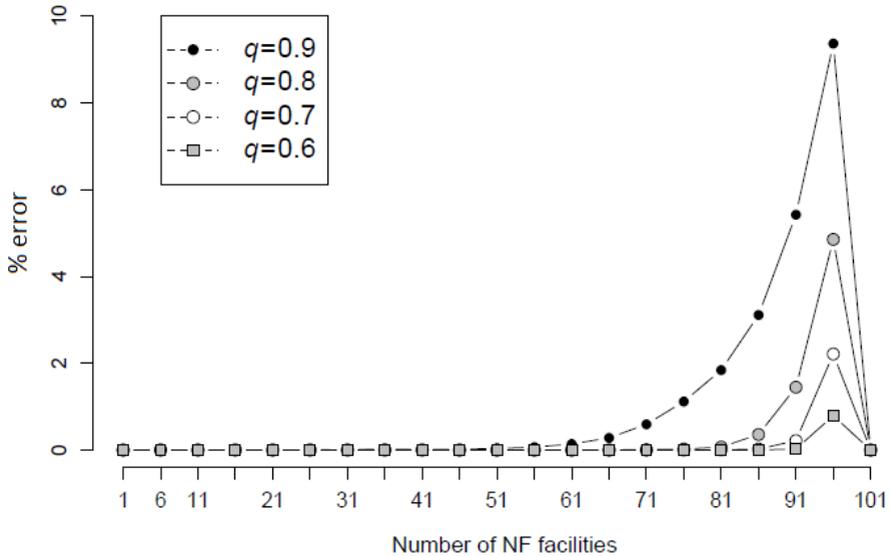

*Figure 1.*   Objective value errors

# A variable neighborhood search approach for the 3-maneuver aircraft conflict resolution problem [*]

A. Alonso-Ayuso,[1] L.F. Escudero,[1] F.J. Martín-Campo,[2]
N. Mladenović[3]

[1]*Universidad Rey Juan Carlos, Madrid, Spain*,  antonio.alonso@urjc.es, laureano.escudero@urjc.es

[2]*Universidad Complutense de Madrid, Madrid, Spain*,  javier.martin.campo@ccee.ucm.es

[3]*Université de Valenciennes et du Hainaut-Cambrésis, France*
Nenad.Mladenovic@univ-valenciennes.fr

## 1. Problem description and resolution

The aircraft conflict resolution problem takes an important role within the Air Traffic Management problem. The aim of the problem consists of providing a new aircraft configuration, almost in real-time, in such a way that every conflict situation during the cruise phase is avoided. A conflict is defined as that event in which two or more aircraft violate the safety distances that must be kept in flight. The distances are 5 nautical miles as horizontal and 2000 feet as vertical. This means that every aircraft is in the center of a safety zone which is a cylinder with 2.5 nautical miles of radius and 2000 feet of height. In order to avoid conflict situations, three maneuvers may be performed: Velocity and heading angle changes as horizontal maneuvers and altitude level changes as vertical ones.

The most critical point in the study of this problem is the resolution time, being necessary to have a solution in almost real-time. It is worth pointing out that the problem to be solved assumes only the aircraft of a certain region of the airspace so-named air sector which may contain up to 20-25 aircraft at most. Our previous studies on the problem are based on a mixed integer nonlinear and nonconvex optimization. The three maneuvers are considered at once. However, the resolution time grows as the number of aircraft under consideration does (increasing the model dimensions), due to the high number of nonlinearities in the constraint system. A metaheuristic approach based on sequentially solving mixed integer linear optimization problems has been also studied but the resolution time is not good when more than 20 aircraft are considered at once. Those models are studied in a multicriteria framework where different criteria are taken into account (lexicographical, compromise, etc.).

[*]This research is partially supported by the projects OPTIMOS3 MTM2012-36163-C06-06 and PCDASO MTM2012-31514, both funded by the Spanish Ministry of Economic Affairs and Competitiveness.

We present a Variable Neighborhood Search (VNS) metaheuristic approach able to provide good solutions by using a small computing time considering the three maneuvers and in a multicriteria environment. A previous work, only considering heading angle changes, reported good results in terms of computing time and solutions quality.

The VNS metaheuristic consists of two important steps: the local search and the shaking phase. The link between them is the penalty cost function composed by two terms in our case: the objective function and the infeasibility condition which is highly penalized. In our case, using a geometric construction presented in a previous work in the literature, the infeasibility condition for horizontal conflict situations is well defined.

For the local search, we have experimented two options: first improvement and best improvement. The previous study with only heading angle changes showed that no important differences on solutions quality appear, so, as the first improvement method needs less computing time, we choose that option. The criterion to go from one solution to another consists of changing the angle for each aircraft until no improvement is found. In the three-maneuvers case, a similar procedure will be implemented, but considering the different maneuvers.

The shaking phase is important to jump from one possible local optimal solution to another zone in which the best solution found can be improved. A subset of aircraft under consideration is selected to change the heading angle. For the three-maneuvers case, the selected aircraft will be allowed to change at least one of the three maneuvers. The new approach combines local search and shaking procedure until the time limit is reached.

A broad computational experience is reported for the case where only heading angle changes are considered, comparing the VNS approach with the state-of-the-art Mixed Integer Nonlinear Optimization engine solver Minotaur and the metaheuristic based on sequentially solving mixed integer linear optimization models. Preliminary results are also presented on the three-maneuvers case by using VNS.

# The probabilistic pickup and delivery problem

E. Benavent[1], M. Landete[2], J.J. Salazar[3], G. Tirado[4]

[1]*Departament d'Estadistica i Investigacio Operativa, Universitat de Valencia, Spain*

[2]*Universidad Miguel Hernandez de Elche, Alicante, Spain*

[3]*Universidad de La Laguna, Spain*

[4]*Universidad Complutense de Madrid, Spain*

In this paper we introduce the Probabilistic Pickup-and-Delivery Travelling Salesman Problem (PPDTSP). It is a variant of the PTSP where one location (called depot) is the starting and ending location of the route, and the others locations are grouped in pairs. Each pair of locations represents a potential request to transport a product from one location (pickup) to the other (delivery). Each potential request may be formalized or not according to a Bernoulli probability distribution. If a potential request is formalized, its pickup location must precede its delivery location along the route. Otherwise, the vehicle must skip visiting the pair of locations associated to that request. A feasible solution for the PPDTSP (also called a-priori route) is a Hamiltonian route to allow serving all requests.Once an a-priori route has being decided, the realization of the random variables is known and the vehicle visits the locations of the formalized requests as they appear in the a-priory route. We present two different formulations and we compare them.

# Multisource linear regression

Víctor Blanco[†] and Justo Puerto[‡]

[†]*Dpt. Quant. Methods for Economics & Business, Universidad de Granada* vblanco@ugr.es

[‡]*Dpt. Statistics & OR, Universidad de Sevilla* puerto@us.es

We present here a mathematical programming approach to linear regression when structural changes occur in the data. We extend the family of methods introduced in [1] where different distance-based residuals and generalized ordered weighted averaging operators to aggregate the residuals were considered. This natural extension is analogous to the one from single-facility location to the multifacility case. General formulations are provided and exhaustive analysis of some especial cases of residuals are reported.

## 1. Introduction

*Regression analysis* studies the functional relationship between a set of variables $X_1, \ldots, X_d$. Such a dependence is expressed through an equation of the form $f(X_1, \ldots, X_d) = 0$, and a sample of data is used to estimate $f$. It is usual to fix the parametric family of functions where $f$ belongs and the parameters are estimated from the sample. In linear regression, $f$ is assumed to be a linear functional and estimating it consists of determining its coefficients.

Given set of data $\{x_1, \ldots, x_n\} \subset \mathbb{R}^d$, to obtain an estimation of the coefficients of the linear model, one tries to find the coefficients of $f$, $\hat{\boldsymbol{\beta}} \in \mathbb{R}^{d+1}$, that minimize some measure of the deviation of the data with respect to the fitting body $\mathcal{H}(\hat{\boldsymbol{\beta}}) = \{z \in \mathbb{R}^d : \hat{\beta}_0 + \sum_{k=1}^d \hat{\beta}_k z_k = 0\}$. For a certain observation $x \in \mathbb{R}^d$ in the sample, such a deviation is known as the *residual*. We consider that the residual of a model is a mapping $\varepsilon_x : \mathbb{R}^{d+1} \to \mathbb{R}_+$, that maps any set of coefficients $\boldsymbol{\beta}$, into a measure that represents how much deviates the fitting of the model, with those parameters, from the observation $x$. The larger this measure, the worse the fitting for such a single ob-servation $x$. The final goal of a regression model is to find the coefficients minimizing a globalizing function, $\Phi : \mathbb{R}^n \to \mathbb{R}$, of the residuals of all the points. Different choices for the residuals and the globalizing criteria will give, in general, different estimations for the parameters and thus different properties for the resulting models. This statistical problem is closely related with continuous location problem, whenever one wants to compute a point (coefficients of the models) provided that it minimizes certain function of the "distances" from the point to a given set of points (sample).

In this paper linear regression models where possible structural changes occur in the data are analyzed. These models are particularly useful when the variables are time-indexed and certain past situations (breakpoints) provoked that the linear relationship of the variables before and after those moments changes. These linear regression models are known in the literature as piecewise regression [5], segmented regression [4] or partition regression.Whenever the breakpoints are known one may apply linear regression methods to all the partitions obtained from the points, reducing to several independent linear regression models, but in practical applications, those points are unknown and need to be determined.

We propose a new family of methods to estimate the coefficients of this structural change problems in linear regression. These methods allows not only to estimate the models but also the breakpoints where the behaviour of the data changes. We point out the analogy between linear regression ($0$ breakpoints problem) and this problem, which is the same that the one between single facility and multi-facility location problems [2, 3].

# A dispersion model in ODEs

Rafael Blanquero [1], Emilio Carrizosa[1], M. Asunción Jiménez-Cordero[1], and Boglárka G.-Tóth[2]

[1]*University of Seville, Spain,*  rblanquero@us.es, ecarrizosa@us.es, asuncionjc@us.es

[2] *Budapest University of Technology and Economics , Hungary,*  bog@math.bme.hu

We consider initial value problems in Ordinary Differential Equations (ODE) in which the initial conditions are decision variables.

Locating the initial conditions yielding $n$ trajectories as disperse as possible (in a sense to be formalised) is a continuous location problem which can be numerically addressed using a geometrical branch and bound.

*Keywords:* dispersion models, ordinary differential equations, location of initial values.

# Rapid transit network design: competition and transfers

Luis Cadarso,[1] and Ángel Marín[2]

[1] *Rey Juan Carlos University, Departamental III Building, Camino del Molino s/n, Fuenlabrada 28943, Spain,*  luis.cadarso@urjc.es

[2] *Universidad Politécnica de Madrid, E.T.S.I. Aeronáuticos, Pza. Cardenal Cisneros, 3, Madrid 28040, Spain,*  angel.marin@upm.es

The problem of increasing traffic congestion is raising the concerns about energy constraints and greenhouse emissions. Considerable attention has been given to the utilization of freight and passengers railway systems as a relative efficient and eco-friendly traffic mode. Increasing mobility and longer journeys caused by the growth of cities have also stimulated the construction and expansion of rail transit systems such as metro, urban rail, and light rail systems. Due to their high cost of construction, it is important to pay close attention to the effectiveness in solving the network design problem. The location decisions and the maximum coverage of the demand for the new network is the main goal, taking a list of potential rapid transit corridors and stations, and considering budget availability.

Transportation demand models are used to develop forecasts of passenger demand for each origin-destination pair as a function of attributes such as average fares, service frequencies, and travel times. Given these total demand estimates, passenger choice models are used to estimate for each competitor and each origin-destination pair, the proportion of demand (or the share) it captures in that origin-destination pair.

Travel time is one of the most important attributes on which the different operators may compete in rapid transit networks. A network can attract more passengers in a origin-destination pair by decreasing travel time. For a given unconstrained total demand, the market share of each operator depends, among other factors, on its own travel time and on the travel time of its competitors. Another convenient attribute for attracting passengers is to offer direct trips without transfers. Transfers are usually extremely discouraging. Both, travel time and transfers strongly depend on the network design. Therefore, it is extremely important to account for them in order to make an efficient design.

In this paper, we present a mixed integer, non-linear programming model for the rapid transit network design problem that includes a passenger choice model to capture multi-modal competition between the new network to be constructed and the current network.

Our major contributions include:

1. Development of a new formulation for the rapid transit network design problem which embeds a logit model for modeling passenger behavior.

2. Introduction of transfers in the modeling approach; a decisive attribute for attracting passengers to the new network is to offer direct trips without transfers. We embed the transfer cost in the passenger trip generalized cost.

3. We solve this model using realistic problem instances obtained from the network of Seville in Spain. Because current commercial software fails to solve medium and large scale problems, we use the Lagrangian Relaxation to obtain the optimal solution of the problem.

# The maximal covering location bi-level problem[*]

José-Fernando Camacho-Vallejo,[1] Martha-Selene Casas-Ramírez,[1] Juan A. Díaz [2] and Dolores E. Luna[2]

[1]*Facultad de Ciencias Físico-Matemáticas, Universidad Autónoma de Nuevo León, Nuevo León, México,*  {jose.camachovl}, {martha.casasrm}@uanl.edu.mx

[2]*Universidad de las Américas Puebla, Puebla, México,*  {juana.diaz}, {dolorese.luna}@udlap.mx

In this presentation the maximal covering location bi-level problem is proposed. The aim of the problem is to maximize the customer's demand covered by a predefined number of opened facilities. The main assumption considered through this research is that customers are allowed to freely choose their allocation to the facilities within a coverage set. The problem is modelled as a bi-level mathematical programming problem where two decision makers are considered: the leader and the follower. The leader is on charge of locate the facilities seeking to maximize the covered demand. On the other hand, the follower will allocate customers to the opened facilities based on their preferences towards the facilities. By doing this, the follower will maximize a utility function that relates those preferences. The problem is optimally solved by two single-level reformulations of the bi-level problem and some equivalency results between the models are discussed. The complexity resulting from preliminary computational experimentation after solving large instances of the problem led us to develop heuristic methods for obtaining good quality solutions. In order to achieve the latter, we implemented a tabu search method and a genetic algorithm. The performance of both heuristics over a large set of instances is shown. Also, the main differences and advantages arising from the numerical experimentation are discussed.

[*]Thanks to the Mexican National Council for Science and Technology (CONACYT) through grant SEP-CONACYT CB-2014-01-240814.

# 1. Problem statement

In this section the problem's description of the maximal covering location bi-level problem (MCLBP) is described. Consider the situation when a set of customers has a demand associated to each of them. These customers are covered -or not- by a existing firm in the market. Then, a competing firm intends to enter to the market by opening some facilities. Therefore, the already covered customers will decide if they stay with they current facility or will change to a new one. Also, the non-covered customers will choose a new facility for being allocated.

The problem is conformed by a leader -the entering competing firm- and a follower -the set of customers-. On the one hand, the leader will decide the location of a limited number of facilities aiming to maximize the demand covered. On the other hand, the follower will allocate the customers to their most preferred opened facilities seeking to maximize the customer's preferences towards the facilities.

Some considerations need to be made by the follower. First, the customers may be allocated to old or new opened facilities; also, some of them may be uncovered. Second, the allocation of the customers can be only made to a facility within a coverage set. That is, if a customer prefers a facility that exceeds the maximum coverage distance, then, the customer cannot be allocated to that facility; instead of, it will be allocated to it most preferred opened facility in the set of facilities that cover that specific customer. Third, the preferences of the customers are assumed to be integer positive consecutive numbers from 1 to the total number of facilities in the market, including the already existing and the potential ones. A greater value implies better preference from that customer to the facility.

Also, it is important to mention that the already existing facilities in the market cannot be closed. It is evident that the leader does not has control over that decision; but, he will try to capture the maximum customer's demand belonging to the existing firm.

# A decomposition scheme for the railway rapid transit depot location and rolling stock circulation problem

David Canca,[1] Alicia de los Santos,[2] Eva Barrena,[3] and Gilbert Laporte[4]

[1]*School of Engineering, Department of Industrial Engineering and Management Science, Av. de los Descubrimientos sn, 41092, Seville, Spain*,   dco@us.es

[2]*School of Engineering,Department of Applied Mathematics II, Av. de los Descubrimientos sn, 41092, Seville, Spain*,   aliciasantos@us.es

[3]*School of Computing,University of Leeds, Leeds, LS2 9JT, United Kingdom*, e.barrena@leeds.ac.uk

[4]*CIRRELT and HEC Montréal, 3000 chemin de la Côte-Sainte-Catherine, Montréal, Canada H3T 2A7*,   gilbert.laporte@cirrelt.ca

Network design and line planning are the two first stages in the railway planning process. However, the network configuration does not include the location of depots where trains have to rest every day. In Rapid Transit Networks, the depot location drastically influences the number and length of empty rolling stock runs, devising in long term high operating costs. Rolling stock is one of the most important operational issues for railway transportation companies. Rolling stock and infrastructure maintenance suppose about 75% of total annual cost for a typical railway transportation. Basically, rolling stock circulation consists of determining individual train paths over the network with the objective of minimizing costs while accomplishing a pre-defined schedule. In this work we formulate the integrated depot location and rolling stock circulation problem and propose a decomposition mechanism in order to solve medium and large instances in reasonable computation times with commercial state-of-the art solvers.

# 1. Introduction

Rolling stock management supposes huge capital investment decisions for service operators that cannot be changed frequently, which means that rolling stock becomes a strategic decision with a future impact of several decades. In general, rolling stock circulation plan includes a set of interrelated sub-problems such as train composition determination (locomotives and carriages coupling and decoupling), vehicle and carriage rest location (avoiding empty movements), vehicle circulation problem and maintenance policies. In the context of RRT systems, services are usually performed by train units with various composition types (e.g. with two, three or four carriages), what means that coupling and decoupling decisions can be avoided. Then, the problem can be viewed as a special multi-commodity capacitated minimum cost flow problem where a set of different commodities (trains with different characteristics) must be routed every day through a network from certain stations (rest places) in order to ensure a set of services and to guarantee minimum operation cost. The problem becomes more complex when train maintenance decisions are also incorporated.

Several models have been proposed for railway rolling stock problems. [11] minimizes the number of train units needed to satisfy a given demand. Related with this paper, [3] study the problem of routing single carriages through a network, considering empty carriage movements. [4] present a Benders decomposition approach for determining a set of minimum cost equipment cycles such that every trip is covered using appropriate equipment. [5] extend their model incorporating maintenance issues. [8] describe a model and a solution approach for a car assignment problem that arises when individual car routings must be determined considering maintenance constraints and minimum connection depending on the positions of cars. [1] present an integer programming model with the objective of minimizing seat shortage during morning rush hours.

[2] propose an integer programming model in order to obtain the circulation of rolling stock considering order in train compositions. The model is devoted to a single line and for only one day of operation. [10] use a transition graph (see [2]) to represent the possible changes in trains composition at each station.

Maintenance issues have also been incorporated in some contributions. [12] set up a large scale integer programming model for locomotive assignment considering that locomotives requiring inspection must be sent to appropriate shops within a given time limit. [5] propose a multicommodity network flow-based model for assigning locomotives and cars to trains. The work of [8] supports operational aspects concerning locomotive-hauled railway cars. They consider maintenance constraints inside a model with the objective of maximizing the expected profit. [9] propose a multicommodity flow type model for routing units that require maintenance in the forthcoming one to three days.

We propose a sequence of optimization models taking advantage of the problem structure in order to first determine the minimum number of vehicles needed to perform the actual train schedule. In a second phase, we obtain the cyclic weekly train circulation which is used to define the most convenient maintenance policy, developed in a third stage.

# Robust $p$-median problem with vector autoregressive demands

Emilio Carrizosa,[1] Alba V. Olivares-Nadal,[1] and Pepa Ramírez-Cobo[2]

[1]*Universidad de Sevilla*

[2]*Universidad de Cádiz*

Most robust location problems are based in either distributional assumptions over the clients' demands, or in scenario analysis. However, lately a new approach is raising: nominal values for the future demands are assumed to be given and uncertainty structures are built taking into account those values. Although authors like to think no statistical procedures are involved in these approaches, the truth is that those nominal values for the demand must have been obtained somehow relying in prediction or estimation methods, and the user has no control over it. Moreover, these uncertainty sets allow demand realizations for time $t$ that do not depend on the realization of the demand for time $t-1$; i.e., these uncertainty sets might contain paths that do not preserve the inner behaviour of the demands.

In this work we assume the demands of the clients follow a Vector Autoregressive Process (VAR), whose parameters are given. Instead of using the nominal values to model the uncertainty sets, we make use of parameters of the VAR. In particular, we aim to solve a robust $p$-median problem where the demand is uncertain but required to follow the VAR process, whose errors are bounded $\left\lVert \right\rVert_p \leq \delta$.

# Using an interior-point method for huge capacitated multiperiod facility location [*]

Jordi Castro,[1] Stefano Nasini,[2] and Francisco Saldanha-da-Gama [3]

[1]*Dept. of Statistics and Operations Research, Universitat Politècnica de Catalunya, Jordi Girona 1–3, 08034 Barcelona, Catalonia, Spain.*   jordi.castro@upc.edu

[2]*Dept. of Production, Technology and Operations Management, IESE Business School, University of Navarra, Av. Pearson 21, 08034 Barcelona, Catalonia, Spain.*   snasini@iese.edu

[3]*Dept. of Statistics and Operations Research/Operations Research Center, Faculdade de Ciências, Universidade de Lisboa, Campo Grande 1749-016 Lisboa, Portugal.*   faconceicao@fc.ul.pt

The capacitated facility location problem is a well known MILP, which has been efficiently solved, among others, by cutting-plane procedures. In this work we focus on a multiperiod variant of this problem. We refer the reader to [4] for a recent survey on facility location.

The solution of the multiperiod capacitated facility location by cutting planes requires the solution of a master problem involving a convex non-differentiable function $Q(y)$, $y$ being the binary variables. This function $Q(y)$ is lower approximated by cutting planes, which are obtained by computing $Q(\hat{y})$ and a subdifferential $s \in \partial Q(\hat{y})$ for some particular $\hat{y}$: this means the solution of a linear optimization subproblem in the continuous variables. In essence, this is Benders decomposition.

Our focus is on world-wide facility locations problems that may be faced, for instance, by internet-based retailer multinational companies. Such problems may involve a few hundreds of world locations for warehousing activities, and millions of customers, associated to cities over some threshold population. In those situations the dimension of the subproblems of the cutting-plane procedure can be of order of hundreds of millions or billions of variables. These subproblems, one for time period, exhibit a block-angular structure that can be exploited by the solver BlockIP [1], which implements a specialized interior-point method for block-angular structures (see, for instance, [2] for details).

---
[*]Work supported by grant MTM2012-31440 of the Spanish Ministry of Economy and Competitiveness

Extensive computational testing showed the efficiency of this combination of cutting planes with interior-point methods for huge problems: the largest instances involved up to 3 time periods, 200 locations and 1 million customers, resulting in MILPs of 600 binary and 600 million continuous variables. Those MILP instances were optimally solved by the cutting-plane-interior-point approach in less than one hour of CPU, while state-of-the-art solvers (i.e., CPLEX) were unable to solve them (either using its internal branch-and-cut MILP solver, or their continuous solvers within the cutting-plane approach). Indeed, the state-of-the-art solver was not able to perform any single iteration since it exhausted the 144 Gigabytes of memory of the server used for these experiments.

# On the hierarchical rural postman problem on a mixed graph

Marco Colombi,[1] Ángel Corberán,[2] Renata Mansini,[1] Isaac Plana,[2] and José M. Sanchis[3]

[1]*Universita degli Studi di Brescia, Italy*,   marco.colombi, renata.mansini@unibs.it

[2]*Universidad de Valencia, Spain*,   angel.corberan, isaac.plana@uv.es

[3]*Universidad Politécnica de Valencia, Spain*,   jmsanchis@mat.upv.es

In this talk we present a special case of the Rural Postman Problem on a mixed graph where arcs and edges that require a service are divided into subsets that have to be serviced in a hierarchical order. In this paper we propose a new mathematical formulation, provide a polyhedral analysis of the problem and propose a branch-and-cut algorithm for its solution based on the introduced classes of valid inequalities.

## 1.    The HMRPP

In many practical applications (garbage collection, snow plowing), constraints may be imposed on the order in which clusters of streets (edges or arcs) have to be visited. For example, in snow plowing, main streets have to be cleared before secondary streets and, in turn, secondary streets should be cleaned before residential ones. Up to now the problem has been analyzed mainly on undirected graphs in which all the edges require a service (the Hierarchical Chinese Postman Problem – HCPP).

In this paper, we extend the HCPP in two ways by first considering mixed instead of undirected graphs and then by requiring a service only on a subset of arcs and edges. The resulting problem, the Hierarchical Mixed Rural Postman Problem (HMRPP), is defined as follows. Let $G = (V, E, A)$ be a strongly connected mixed graph, where $V$ is the set of nodes, $E$ is the set of edges, and $A$ is the set of arcs. Node 1 is the depot. From now on, we will use the term link to refer both to an edge or an arc, indistinctly.

Links requiring a service are divided into classes (hierarchies) indicating a predefined priority on the service.

Let $E_R = E_R^1 \cup E_R^2 \cup ... \cup E_R^p$ be the set of required edges and $A_R = A_R^1 \cup A_R^2 \cup ... \cup A_R^p$ the set of required arcs, where $p$ is the number of hierarchies. Following the specified hierarchical order, all the links in $E_R^k \cup A_R^k$ must be serviced before those in $E_R^m \cup A_R^m$ if $k < m$.

As in [1], we consider three different costs associated with each required link $(i, j) \in E_R \cup A_R$:

- $\hat{c}_{ij}$ : the cost of traversing a link $(i, j)$ that has not been serviced yet,
- $\bar{c}_{ij}$ : the cost of traversing and servicing link $(i, j)$,
- $\tilde{c}_{ij}$ : the cost of traversing a link $(i, j)$ that has already been serviced.

A crossing cost $c_{ij}$ is associated with each non-required link $(i, j)$. The cost of traversing an edge is the same in both directions. Usually it is assumed (see [1]) that $\bar{c}_{ij} > \hat{c}_{ij} > \tilde{c}_{ij}$. Consider, for example, snow removal operations, where removing snow on a street is the most time-consuming operation, followed by traversing an uncleaned street and, finally, crossing an already cleaned one. Nevertheless, other sorting of costs may find a practical justification as well. For instance, it is not difficult to assume that in snow removal $\hat{c}_{ij} > \bar{c}_{ij}$ holds for some links, that is, the cost of traversing a non serviced link is definitely more expensive than that of servicing it since a non-cleaned street can be hardly traversed.

The HMRPP consists of finding a minimum cost tour starting and ending at the depot while servicing all the required links in the specified hierarchical order. Note that the subgraph induced by the required links $E_R \cup A_R$ and the subgraphs induced by the sets $E_R^k \cup A_R^k$, for all $k$, do not need to be connected. Moreover, since the required links must be serviced in a hierarchical order, a tour for the HMRPP can be interpreted as a sequence of consecutive connected paths, each one servicing the required links in each hierarchical level, plus a final path back to the depot.

In this paper we propose a new mathematical formulation, provide a polyhedral analysis of the problem and propose a branch-and-cut algorithm for its solution based on the introduced classes of valid inequalities.

# Strategic oscillation for a hub location problem with modular link capacities

Ángel Corberán[1], Juanjo Peiró[1], Fred Glover[2], and Rafael Martí[1]

[1]*Departament d'Estadística i Investigació Operativa, Universitat de València, Spain.*
angel.corberan@uv.es; juanjo.peiro@uv.es; rafael.marti@uv.es

[2]*OptTek Systems, Boulder (CO), USA.*
glover@opttek.com

Hub location problems related to the design of transportation networks are one of the most extensively studied problems in combinatorial optimization due to their variety and importance. In a hub location problem, one is given a network $G = (V, E)$ with a set of demand nodes $V$, and a set of edges $E$. For each pair of nodes $i$ and $j \in V$, there is a traffic $t_{ij}$ (of goods, people, etc.) to be transported through a set of facility locations selected from a given set of potential locations $V$. It is assumed that direct transportation between terminals is not possible and, therefore, the traffic $t_{ij}$ travels along a path $i \to h_i \to h_j \to j$, where $i$ and $j$ are assigned to hubs $h_i$ and $h_j$, respectively. The goal is to identify an optimal subset of facilities in order to minimize a transportation cost function while satisfying a set of constraints.

Among the family of hub location problems, we focus on a specific variant known as the *capacitated single assignment hub location problem with modular link capacities* (CSHLPMLC). This problem was formulated as a quadratic mixed integer programming problem by Yaman and Carello [1], with the following characteristics: Opening a hub at node $i$ has a fixed installation cost $C_{ii}$. Each hub $i$ has a capacity $Q^h$ limiting the total amount of traffic transiting through $i$. There are two types of edges between nodes: edges of the first type are used to connect terminals with hubs, and we call them *access edges*. Let $m_i$ be the number of access edges needed to route the incoming and outgoing traffic at node $i$, and let $Q^a$ be the maximum capacity an access edge allows to transfer through it in each direction. So,

$$m_i = \max\left\{\left\lceil\frac{\sum_{j\in V} t_{ij}}{Q^a}\right\rceil, \left\lceil\frac{\sum_{j\in V} t_{ji}}{Q^a}\right\rceil\right\}.$$ The cost of installing $m_i$ edges between terminal $i$ and hub $k$ is denoted by $C_{ik}$. Edges of the second type are used to transfer traffics between hubs, and we call them *backbone edges*. Each backbone edge has a maximum traffic capacity of $Q^b$ (in each direction). We define $A$ as the set of (directed) arcs associated with the (undirected) edges in $E$, $A = \{(k,l) : k, l \in V, k \neq l\}$.

If nodes $k$ and $l$ are hubs, the amount of traffic on arc $(k,l)$, denoted as $z_{kl}$, is the traffic that has to be transported from nodes assigned to $k$ to nodes assigned to $l$. The capacity $Q^b$ of a given edge $\{k,l\}$ cannot be less than the maximum traffic on its corresponding arcs $(k,l)$ and $(l,k)$, and the cost of installing the edge is denoted by $R_{kl}$.

Three different costs have to be considered in this problem: The opening costs of the hubs ($C_{kk}$), the assignment costs of terminals to hubs ($C_{ik}$), and the traffic costs between hubs. While cost $C_{ik}$ corresponds to that of transporting all the traffic involving $i$ through hub $k$, $R_{kl}$ represents the cost of using only one backbone edge $\{k,l\}$.

The following variables are defined in [1] in order to provide the mathematical programming model shown below:

- The assignment variable $x_{ik}$ is equal to 1 if terminal $i$ is assigned to hub $k$, and 0 otherwise. If node $i$ receives a hub, then $x_{ii}$ takes value 1.

- $z_{kl}$ is the traffic on an arc $(k,l) \in A$ and $w_{kl}$ is the number of copies of the edge $\{k,l\} \in E$.

Then, the capacitated single assignment hub location problem with modular link capacities can be formulated as follows ([1]):

$$\min \sum_{i\in V}\sum_{k\in V} C_{ik}x_{ik} + \sum_{\{k,l\}\in E} R_{kl}w_{kl} \quad (1)$$

$$\sum_{k\in V} x_{ik} = 1 \quad \forall i \in V \quad (2)$$

$$x_{ik} \leq x_{kk} \quad \forall i \in V, \quad \forall k \in V \setminus \{i\} \quad (3)$$

$$\sum_{i\in V}\sum_{j\in V}(t_{ij}+t_{ji})x_{ik} - \sum_{i\in V}\sum_{j\in V} t_{ij}x_{ik}x_{jk} \leq Q^h x_{kk} \quad \forall k \in V \quad (4)$$

$$z_{kl} \geq \sum_{i\in V}\sum_{j\in V} t_{ij}x_{ik}x_{jl} \quad \forall (k,l) \in A \quad (5)$$

$$Q^b w_{kl} \geq z_{kl} \quad \forall \{k,l\} \in E \quad (6)$$

$$Q^b w_{kl} \geq z_{lk} \quad \forall \{k,l\} \in E \quad (7)$$

$$x_{ik} \in \{0,1\} \quad \forall i, k \in V \quad (8)$$

$$w_{kl} \in \mathbb{Z}_+ \quad \forall \{k,l\} \in E \quad (9)$$

$$z_{kl} \geq 0 \quad \forall (k,l) \in A. \quad (10)$$

Many heuristics and metaheuristics have been proposed to solve different variants of hub location problems, including GRASP, VNS, tabu search, and several complex hybrid techniques. In this research we present a simple, easily adaptable and powerful algorithm, based on the iterated greedy – strategic oscillation (SO) methodology. Our purpose is to investigate the SO proposal, which alternates between constructive and destructive phases as a basis for creating a competitive method for this hub location problem. We will also present a comparison between methods and optimal results. Computational outcomes on a large set of instances show that, while only small instances can be optimally solved with exact methods, our metaheuristic is able to find high-quality solutions on larger instances in short computing times, and outperforms the previous tabu search implementation.

# Ordered Weighted Average Optimization in multiobjective spanning tree problems

Elena Fernández,[1] Miguel A. Pozo,[2] Justo Puerto[2] and Andrea Scozzari[3]

[1]*Universitat Politècnica de Catalunya – Barcelona Tech (Spain)*　e.fernandez@upc.edu

[2]*University of Seville (Spain)*　miguelpozo@us.es, puerto@us.es

[3]*University Niccolò Cusano. Roma (Italy)*　andrea.scozzari@unicusano.it

Multiobjective Spanning Tree Problems are analyzed in this paper. The ordered median objective function is used as an averaging operator to aggregate the vector of objective values of feasible solutions. This leads to the study of the Ordered Weighted Average Spanning Tree Problem, a nonlinear combinatorial optimization problem. Different formulations as a mixed integer linear problem are proposed, based on the most relevant Minimum cost Spanning Tree models in the literature. These formulations are analyzed and several enhancements proposed. Their empirical performance is tested over a set of randomly generated benchmark instances. The results of the computational experiments show that the choice of an appropriate formulation allows to solve larger instances with more objectives than those previously solved in the literature.

# Set-packing problems in discrete location

Alfredo Marín[1] and Mercedes Pelegrín[2]

[1]*Murcia University, Murcia, Spain*   amarin@um.es

[2]*Murcia University, Murcia, Spain*   mariamercedes.pelegrin@um.es

Set-packing problems are a typical matter of discrete optimization in general. Given a collection of subsets of $\{1,\ldots,n\}$ together with their weights, set-packing problems consist of optimize a linear objective function satisfying that every pair of subsets in the group has not any point in common. Many practical location issues can be formulated as set-packing problems, for instance the uncapacitated facility location problem, two-stage location problems, hub location problems and so on. Some other times only a part of the formulation can be described as a set-packing problem. This work reviews different approaches that have been used to solve discrete location problems from the scope of the set-packing problem.

# Robust rapid transit railway rescheduling [*]

Ángel Marín,[1] and Luís Cadarso[2]

[1]*Technical University of Madrid, Aeronautic and Space ETSI, Pl. Cardenal Cisneros, 3, 28040 Madrid, Spain,*  angel.marin@upm.es

[2]*Rey Juan Carlos University, Departamental III Building, Office 012, 28943, Fuenlabrada, Madrid, Spain,*  luis.cadarso@urjc.es

A new integrated approach to the Rail Scheduling problem for the rolling stock (RS) assignment, train maintenance routing and crew scheduling problems in rapid transit networks (RTN) is proposed all while considering the passengers use optimization. RS circulation with train routing and integrated short-term maintenance is a large problem, so it is basic consider that maintenance required covering a given set of services and consider maintenance covering a given set of tasks and works with a minimum amount of rolling stock units, avoiding the empty services and covering a given passenger demand.

Well-known examples of these RT planning models are strategic, tactical and operative problems addressed during the planning process. Due to the tremendous size of the planning process, it is usually divided into several steps such as:

1. Network design: the infrastructure location decisions and the maximum coverage of the demand for the new network are the main goal here. ([5]).

2. Line planning: designing a line system (i.e., determining origin and destination stations, stops and frequencies) aims at satisfying the travel demand while maximizing the service towards the passengers or minimizing the operating costs. ([4]).

3. Timetabling: The general aim of the railway timetabling problem is to construct a train schedule that matches the frequencies determined in the line planning problem.([2]).

[*] This research was supported by Project Grants TRA2011-27791-C03-01 and TRA2014-52530-C3-1-P by the Ministerio de Economía y Competitividad, Spain.

4. Given a fleet of train units and finding the optimal train unit assignment to each train to satisfy both the timetable and the demand in a dense RTN is known as the RT RS assignment problem. The routing problem is the process of determining a sequence for each train unit in the network once the RS assignment is known. These routing must verify the maintenance constraints, the goal is to obtain rolling sequences that minimize some cost such as the propagated delay in order to achieve a robust solution.([1]).

5. The overall crew management problem is usually approached in two phases, namely crew scheduling and crew rostering. Here, we consider the short-term schedule of the crews. A duty represents a sequence of operations to be covered by a single crew within a given time period (e.g., a work day). The long-term schedule of the crews is out of the scope of this paper. ([3]).

   The simultaneous consideration of the crew assignment joint with the above considerations drive to large models, will require total train service and demand coverage. We solve the integrated model using different decomposition and heuristic methods evaluating their performances. Preliminary computational experiments for real case studies drawn from RENFE (the main Spanish train operator) show that the efficient of the integrated models and methods.

# New products supply chains: the effect of short lifecycles on the supply chain network design.

Mozart B.C. Menezes,[1] Kai Luo,[1] and Oihab Allal-Cherif[1]

[1]*Operations Management and Information Systems Department*
*Research Cluster in Supply Chain Management*
*Kedge Business School, France*

We attempt to shed light on the effect of stochastic demand on the location and capacity of production facilities. We consider the very early stages of a new product development; i.e., locational decisions about warehouse and production facilities are being made as well as their capacities, outsourcing decisions are also being made through capacity reservation, markets are tested for getting knowledge about potential demand. At that time, there is the stochastic total demand $D$ where each demand point $i$ in set $N$ will get a fraction $\lambda_i$ of the total demand $D$. Thus, we consider that there is full correlation on the realization of demand seen by each demand point. This is the case when the major source of demand variability is the state of the world in terms of economic situation.

The framework is that of a traditional Newsvendor problem where decisions will generate expected under- and over-capacity costs, which are function of both unitary cost of acquiring capacity and transportation cost (function of facility location). What makes this work different from several other inventory-location problems is that in this work the 'critical fractile' is not uniform across facilities. We focus on the situation where decision makers may attempt to provide higher service level for a group of demand points and lower service levels to another. The objective of the decision maker is that of profit maximization after computing the revenue, the transportation costs, capacity costs and the cost of not being able to serve demand from the capacity in place. The major constraint for the decision maker is that the *return-on-assets* is above a pre-stipulated threshold.

This paper fits the stream of research aimed at jointly considering facility location and inventory management. On the facility location side, It is as the generalization of the classical Capacitated Facility Location Problem (CFLP) which seeks to find optimal location for a certain number of new facilities so as to maximize the amount of demand captured. It can also be viewed as a generalization of the maximization with fixed costs version of the $p$-median model, Hakimi [2].

Effects of inventory and location decisions originate with Erlebacher and Meller [1] who introduced a highly non-linear integer model that combines location and inventory decisions. Shen [3] and Shen, Coullard and Daskin [5] develop several efficient algorithms for the joint inventory location problem that considers costs of facilities, shipments and safety stocks. The algorithms include Branch and Price and Lagrangian relaxation methods. For more on inventory location models, readers can refer to the recent survey paper of Shen [4].

We depart from the previous work by considering that there is no uniform service level (as most work above assumes) which makes the problem more difficult. We also do not consider that each facility has a fixed partition of the network to serve. On the other hand, by assuming perfect correlation of node-demand, we simplify the problem and make it more tractable. Note that this last assumption is perfectly aligned with the framework of the problem and, hence, does not constitute a strong assumption. Finally, we shift the focus of our work to structural properties and the insights they can bring as opposed to numerical approaches where previous work have focused on.

# Dial-a-ride problems in presence of transshipments

Juan A. Mesa,[1] Francisco A. Ortega[2] and Miguel A. Pozo[3]

[1]*Dep. of Applied Mathematics II. Higher Tech. School of Engineering, University of Seville, Camino de los Descubrimientos s/n, Seville 41092, Spain,* jmesa@us.es

[2]*Dep. of Applied Mathematics I. Higher Tech. School of Architecture, University of Seville, Av. Reina Mercedes 2, Seville 41012, Spain,* riejos@us.es

[3]*Dep. of Statistics and Operative Research. Faculty of Mathematics, University of Seville, C/ Tarfia s/n, Seville 41012, Spain,* miguelpozo@us.es

The Dial-A-Ride Problem (DARP) consists of determining a number of routes of minimum cost in order to satisfy to a set of transportation requests. Each request implies transporting a group of users from their respective origins, called pickup points to their corresponding destina-tions, called delivery points. Users who belong to different requests can share the same vehicle while its capacity is not exceeded.

For each request, there exists a maximum ride time associated with the maximum duration of the trip between the pickup point and the deliv-ery point. This work addresses the generalization of the DARP when the users can be transferred from one vehicle to another at specific in-termediate points, called transfer points.

# Optimal routes to a destination for airline expansion

B. Pelegrín Pelegrín,[1] P. Fernández Hernández,[1]
and J.D. Pelegrín García[2]

[1]*Department of Statistics and Operational Research, University of Murcia,*
{pelegrin,pfdez}@um.es

[2]*Department of Engineering Science, University of Oxford,*
juan.pelegringarcia@stcatz.ox.ac.uk

**Abstract**

A topic of interest for airline managers is whether the company could increase its revenue by increasing the number of flights to a specific destination. When a sufficient number of travelers in a certain region with multiple airports is expected to travel to the destination, the airline has a chance to maximize its revenue by gaining a higher portion of the market. For this reason managers might be interested in knowing how to maximize market share by making decisions on flight frequency from each airport in the region to the destination. The market share captured by the airline will not only depend on the airline decision, but also on the airlineś competitors and their ability to perform efficiently in the market. Therefore, managers need to know about the airports and the airline choice criteria used by the air travelers in the region in order to estimate the airline market share.

To date, a number of studies have modeled passenger choices of airport and airline by taking into account several factors such as the distance from departure point to airport, ticket price, frequency of flights, timetable, and level of the service. The passenger demand of the airline could vary with respect to the current demand if any of such factors changes (see for instance [1],[2],[5]). Most studies use discrete choice models to predict the probability that a traveler will choose one alternative among a finite set, based on the attributes of the different alternatives which are integrated in a variety of utility functions. The most common models used in airline competition are the multinomial logit (MNL) model ( see [3],[4],[6],[10]) and the nested logit (NL) model ( see [8],[11],[12],[13]). These studies analyze the significant factors affecting passenger choice behaviour by using regression analysis tools for parameter estimation based on survey data collected via mail, telephone and face to face interviews. By using such probabilities, the market share can be expressed in different ways depending on the variables considered. To our knowledge, the effect of increasing flight frequency on market share has not been addressed in the existing literature. This may be due to the fact that the functions obtained for market share estimation by using the MNL and NL models are not of any standard type and then standard optimizers can not be employed to solve the corresponding maximization problem.

In a region with multiple airports many travelers are willing to spend 3 or 4 hours to drive to larger metropolitan airports where they have more choices of airlines and departure times with lower airfares. This behaviour seems to suggest that distance is a key factor in our problem of market share estimation. In this paper, we use a Huff-like (see [7, 9]) spatial interaction model for airline market share estimation together with an utility function which depend on distance to the airport, airport characteristics, ticket price and flight frequency. Then we develop a formulation of the airline market share maximization problem when the decision variables are flight frequencies which can be solved by standard optimizers. An illustrative example with data from Spanish airports is presented which shows that real size problems can be solved by using the proposed formulation.

# Locating new stations and road links on a road-rail network

Federico Perea [1]

[1]*Instituto Tecnológico de Informática.*
*Universitat Politècnica de València (Spain),*  perea@eio.upv.es

In this paper we study the problem of locating new stations on an existing rail corridor and new junctions on an existing road network, and connecting them with a new road segment under a budget constraint, with the objective of attracting as many passengers as possible.

## 1. Introduction

The advantages of railway systems with respect to other transportation modes in short-medium distances are commonly agreed: non-dependence on petrol, reduction of polluting emissions, higher safety, comfort, etc. For this reason, many new railway systems are being constructed from scratch or extended. The latter problem is what we are dealing with in this work. Following previous references, this problem is here called the *railroad network extension problem*. In such problem two different transportation networks, namely a road network and a railway network, are both competing and collaborating. They compete in the sense that each network wants to attract as many travelers as possible. They collaborate in the sense that multi-modal trips, in which both transportation systems are used to reach your destination, are allowed.

## 2. The problem

The problem here presented considers a *road-rail* connected network, which consists of road links and rail links, as well as cities, train stations and junctions. Using graph terminology, we consider an undirected road-rail net-work $G = (R \cup T, E_R \cup E_T)$. As usual, $R \cup T$ is the node set and $E_R \cup E_T$ is the edge set. The node subset $R$ consists of a set of cities and junctions without a train station, and the node subset $T$ consists of cities and junctions with a train station. As for the edges, the elements in $E_R$ are road links directly joining two nodes of $R$, or one node of $R$ with one node of $T$. Finally, the edges in $E_T$ directly link two railway stations, that is, two nodes of $T$.

As an example, consider Figure 1. The existing stations in set $T$ are represented by filled squares. The cities and junctions without a train station in set $R$ are represented by filled circles. The road links $E_R$ are represented by dotted lines, and the railway links $E_T$ are represented by filled lines.

The aim is to build new stations on the rail network (in the figure represented by empty squares) and new junctions on the road network (in the figure represented by empty circles), and to build road links between the new stations and the new junctions (in the figure represented by dashed lines), optimizing a certain objective. The most commonly used objective is the maximization of the ridership: the number of potential travelers that would prefer the railway network over other existing transportation modes.

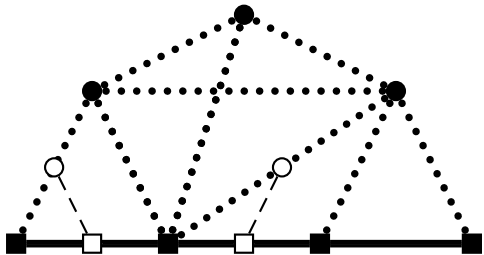

*Figure 1.*  An extended rail-road network.

This research extends [1], where only one station and one junction were located. The models and algorithms there introduced will serve as a starting point to solve this railroad network extension problem.

# Convex ordered median problem

Diego Ponce[1] and Justo Puerto[1]

[1]*Universidad de Sevilla*
dponce@us.es, puerto@us.es

The Ordered Median Problem is a modeling tool that provides flexible representations of a large variety of problems, which include most of the classical location problems considered in the literature. It consists in minimizing a globalizing function that assigns weights depending not of the costs induced by the facilities themselves but to their position in the relative vector of ordered costs. In the discrete version the assignment costs are defined a priori or they are induced from the distances between nodes in a network. In this talk we present specific formulations for the Continuous Convex Ordered Median Problem, which is the continuous version of the above problem that occurs whenever the costs are induced from distances between point in a continuous space and the vector of lambda weights is monotone no decreasing. We exploit new findings to this particular problem, such as the column generation algorithm or recently developed techniques for solving SOC programs .

# On $k$-centrum optimization with applications to the location of extensive facilities on graphs and the like

Justo Puerto,[1] Antonio M. Rodríiguez-Chía,[2]

[1]*IMUS, Universidad de Sevilla, Spain* puerto@us.es

[2]*Universidad de Cádiz, Spain* antonio.rodriguezchia@uca.es

**Keywords:** *Ordered Median Problems, Extensive Location*

This talk addresses a class of combinatorial optimization models that include among others, bottleneck and k-centrum and that extends further to general ordered median objective functions. These problems have been analyzed under different names for different authors in the last years ([1], [2], [3], [4], [5], [6, 7], [8] and [9]). We study the common framework that underlines those models, present different formulations and study some relationships and reinforcements. This approach leads to polynomial time algorithms for the location of extensive facilities on trees that were not previously known as for instance the $k$-centrum subtree and $k$-centrum path location on trees.

# Analyzing the impact of capacity volatility on the design of a supply chain network

Ruiz-Hernández, Diego,[1] Menezes, Mozart B.C.,[2] and Gueye, Serigne[3]

[1]*University College for Financial Studies - Leonardo Prieto Castro 2, 28040, Madrid, Spain,* d.ruiz@cunef.edu

[2]*KEDGE Business School - Bordeaux, France,* Mozart.Menezes@kedgebs.com

[3]*Université d'Avignon et des Pays du Vaucluse - Avignon, France,* serigne.gueye@univ-avignon.fr

Service facilities can and usually, at some point in time, do fail. Failures originate from different sources, ranging from natural disasters and locally disruptive events (e.g. tornados, strikes, and so on), to temporary shortages of capacity due to human or natural factors (e.g. work accidents, electricity outages, and so on). In order to mitigate the financial effect of these capacity disruptions, managers must take decisions aimed at guaranteeing the availability and reliability of their distribution network. Among the aspects that must be taken into consideration we can highlight the capacity and resilience of the facilities (i.e. how resistent they are to exogenous disruptions) and their location.

While the classical location theory tended to ignore the reliability aspects, there is a growing interest in better understanding the impact of imperfect reliability on the optimal location patterns and the resulting costs.

In this work we aim at shedding light on the problem of locating facilities in the knowledge that their respective capacities may be disrupted and, consequently, reduced from their theoretical (planned) value. We examine a class of problems where service facilities may fail, understanding a failure as a deviation of the facility's capacity from its theoretical value, given certain probability distribution. Moreover, we allow for the possibility of failures to be correlated across different facilities.

The problem is parameterized by introducing, for each facility, a couple of values, $C$ and $\beta$, that represent the facility's theoretical capacity and its resilience to external shocks, respectively. Therefore, a manager concerned about availability and reliability of its distribution network may be interested on investing either in capacity (increasing $C$), in resilience (increasing the facility's $\beta$) or in both of them simulataneously.

Moreover, for a given combination of capacity and resilience, the financial impact of a capacity shortcut may depend on both the location of the facility and the location of other sister facilites that may be available for providing backup services. Therefore, strategic location plays an important role for mitigating the economic impact of capacity fluctuations. In order to address this, we propose the *Unreliable Capacitated Facility Location Problem* (UCFLP) that aims at maximizing profit by means of identifying the optimal location of a number of facilities. The problem is formulated as a stochastic programming problem, where a collection of scenarios is used to model the occurrence and severity of exogenous disruptions.

The proposed UCFLP extends the Median Problem with Unreliable Facilitis proposed by Berman et al. [1] in three main directions: (i) it considers correlated failure events; (ii) it does not restrict failures to complete losses of capacity, i.e. a disruption can affect a faciliy's capacity only partially; and (iii) introduces fixed costs, empowering the endogenous election of the number of facilities.

An extensive numerical investigation is conducted to asses the practical implications of our model. In particular, we pay special attention to the effects of diferent configurations of parameters $C$ and $\beta$ on the spacial distribution (concentration/dispersion) of the facilities. We also investigate the financial benefit, for a given location, of strategic increases in the facility's capacity and/or resilience.

# Location of emergency units in collective transportation line networks *

Tunzi Tan,[1] and Juan A. Mesa,[2]

[1]*University of Chinese Academy of Sciences, Beijing, China,*   ttzwin@mails.ucas.ac.cn

[2]*University of Seville, Seville, Spain,*   jmesa@us.es

Due to different causes, disruptions on transportation networks often happen. For example, in railway networks, infrastructure failures (failing signals, broken catenary, etc.), rolling stock failures (engine break-downs, power failures, problems with closing doors, etc.), human factors of the crew (drivers out of conditions), accident with other traffic, suicides and vandalism (stones or coins on the tracks), disturb railway traffic and require a quick response of the operator company. In most cases repairmen should go to the place where those accidents happen and take actions.

A railway system can be represented by a graph where each station is represented by a node and the links between adjacent stations are shown by edges. However, this representation does not capture all the characteristics of a railway network which is structured by lines. For this purpose the concept of hypergraph is useful. In the railway representation by hypergraph, stations are nodes and lines are hyperedges. The combination of both concepts is called hyperstructure and captures both the binary relationship between adjacent stations and the multiple relationship of the stations belonging to the same line.

In this paper, we tackle the problem of locating emergency units in a railway network. For this purpose, we use a minmax hypergraph partition of the network with balanced constraints which allows us to apply in each part a center location problem to determine the stations where to establish the emergency units.

The minmax hypergraph partition problem, which only recently has been defined [1], aims at partition a hypergragh in parts so that the max-

*This research was partially supported by Ministerio de Economía y Competitividad (Spain)/FEDER under grant MTM2012-37048

imum number of hyperedges covering each part is minimized. Balanced constraints are imposed in order to equilibrate the number of stations in each part. Graph partition problem is NP-hard, so is the hypergraph partitioning problem. For this purpose, apart of designing exact algorithms it is also required to design approximation and/or heuristic algorithms. Also a mathematical programming formulation of this problem is provided. The paper finishes with some computational experience and the application to actual metro networks.

# Author Index

Sponsors:

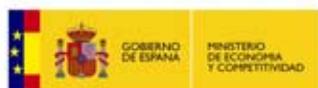
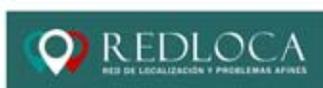
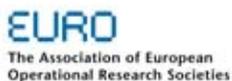

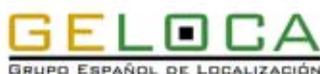
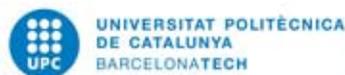
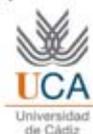

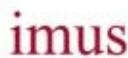